\pdfoutput=1
\documentclass[aps,prd,amsmath,floats,floatfix, twocolumn,
superscriptaddress,nofootinbib,showpacs,longbibliography]{revtex4-1}
\usepackage[T1]{fontenc}
\usepackage[utf8]{inputenc}
\usepackage{lmodern}

\usepackage{verbatim}

\usepackage[dvipsnames, usenames]{xcolor}
\definecolor{linkcolor}{rgb}{0.0,0.3,0.5}
\usepackage[hypertexnames=false, unicode, colorlinks=true, linkcolor=linkcolor,
citecolor=linkcolor, filecolor=linkcolor,urlcolor=linkcolor,
pdfusetitle]{hyperref}

\usepackage[all]{hypcap}
\usepackage{graphicx}
\usepackage{xspace}
\usepackage{amssymb}
\usepackage[normalem]{ulem}
\usepackage{bm}

\usepackage{microtype}

\usepackage{etoolbox}

\usepackage[english]{babel}
\usepackage{blindtext}

\newcommand\prlsec[1]{\vspace{2mm}\noindent \textbf{\emph{#1}}---}

\graphicspath{%
  {figs/}%
}

\DeclareMathAlphabet{\mathpzc}{OT1}{pzc}{m}{it}

\newtoggle{commentsoff}
\togglefalse{commentsoff}

\ifdefined\nocomments
    \toggletrue{commentsoff}
\fi

\iftoggle{commentsoff}{

    \newcommand{\Note}[1]{}
    \newcommand{\TODO}[1]{}
    \newcommand{\AddCite}[1]{}
}{

    \newcommand{\Note}[1]{\textcolor{blue}{\textbf{[#1]}}}
    \newcommand{\TODO}[1]{\red{TODO: #1}}
    \newcommand{\AddCite}{\red{[Needs citation]}}
}

\newcommand{\red}{\textcolor{red}}

\newcommand{\trefmHundredM}{t_{\mathrm{ref}}/M\!=\!-100}

\newcommand{\frefTwentyHz}{f_{\mathrm{ref}}\!=\!20 \, \mathrm{Hz}}

\newcommand{\Jeffreys}{\texttt{Jeffreys}-$\sigma_{\phi}$\xspace}
\newcommand{\Flat}{\texttt{Flat}-$\sigma_{\phi}$\xspace}
\newcommand{\Infinite}{\texttt{Infinite}-$\sigma_{\phi}$\xspace}
\newcommand{\delphi}{\Delta \phi}

\newcommand{\bchi}{\bm{\chi}}
\newcommand{\bL}{\bm{L}}

\newcommand{\Spins}{\mathbb{S}}
\newcommand{\Li}{\mathcal{L}}

\newcommand{\mOneSrc}{m^{\mathrm{src}}_{1}}
\newcommand{\R}{R}
\newcommand{\vesc}{v^{\mathrm{max}}_{\mathrm{esc}}}
\newcommand{\GCRange}{3 - 10}
\newcommand{\NSCRange}{39 - 73}

\newcommand{\PhenomT}{\texttt{IMRPhenomTPHM}\xspace}
\newcommand{\NRSur}{\texttt{NRSur7dq4}\xspace}
\newcommand{\NRSurRemnant}{\texttt{NRSur7dq4Remnant}\xspace}

\newcommand{\spinframefignum}{1\xspace}
\newcommand{\spinpopfignum}{2\xspace}
\newcommand{\kickpopfignum}{3\xspace}
\newcommand{\bayeseqnum}{1\xspace}
\newcommand{\reweightedbayeseqnum}{2\xspace}
\newcommand{\hierbayeseqnum}{3\xspace}
\newcommand{\hierLieqnum}{4\xspace}

\newcommand{\numEv}{31\xspace}

\newcommand{\TITLE}{Hints of spin-orbit resonances in the binary black hole
population}

\newcommand{\Cornell}{\affiliation{Cornell Center for Astrophysics
    and Planetary Science, Cornell University, Ithaca, New York 14853, USA}}
\newcommand\CornellPhys{\affiliation{Department of Physics, Cornell
    University, Ithaca, New York 14853, USA}}

\newcommand{\AEI}{\affiliation{Max Planck Institute for Gravitational Physics
    (Albert Einstein Institute), Am M\"uhlenberg 1, Potsdam 14476, Germany}} %

\newcommand\MIT{\affiliation{LIGO Laboratory, Massachusetts Institute of
Technology, Cambridge, Massachusetts 02139, USA}}
\newcommand{\MKI}{\affiliation{Department of Physics and Kavli Institute for Astrophysics and Space Research, Massachusetts Institute of Technology, 77 Massachusetts Ave, Cambridge, MA 02139, USA}}
\newcommand{\CCA}{\affiliation{Center for Computational Astrophysics, Flatiron Institute, New York NY 10010, USA}}
\newcommand{\StonyBrook}{\affiliation{Department of Physics and Astronomy, Stony Brook University, Stony Brook NY 11794, USA}}

\begin{document}

\title{\TITLE}

\author{Vijay Varma}
\email{vijay.varma@aei.mpg.de}
\thanks{Klarman fellow; Marie Curie fellow}
\CornellPhys
\Cornell
\AEI

\author{Sylvia Biscoveanu}
\MIT
\MKI
\author{Maximiliano Isi}
\thanks{NHFP Einstein fellow}
\MIT
\MKI

\author{Will M. Farr}
\StonyBrook{}
\CCA{}

\author{Salvatore Vitale}
\MIT
\MKI
\hypersetup{pdfauthor={Varma el al.}}

\date{\today}

\begin{abstract}
Binary black hole spin measurements from gravitational wave observations can
reveal the binary's evolutionary history. In particular, the spin orientations
of the component  black holes within the orbital plane, $\phi_1$ and $\phi_2$,
can be used to identify binaries caught in the so-called spin-orbit resonances.
In a companion paper, we demonstrate that $\phi_1$ and $\phi_2$ are best
measured near the merger of the two black holes.
In this work, we use these spin measurements to provide the first constraints
on the full six-dimensional spin distribution of merging binary black holes.
In particular, we find that there is a preference for $\Delta \phi = \phi_1 -
\phi_2 \sim \pm \pi$ in the population, which can be a signature of spin-orbit
resonances. We also find a preference for $\phi_1 \sim -\pi/4$ with respect to
the line of separation near merger, which has not been predicted for any
astrophysical formation channel.  However, the strength of these preferences
depends on our prior choices, and we are unable to constrain the widths of the
$\phi_1$ and $\Delta \phi$ distributions. Therefore, more observations are
necessary to confirm the features we find. Finally, we derive constraints on
the distribution of recoil kicks in the population, and use this to estimate
the fraction of merger remnants retained by globular and nuclear star clusters.
We make our spin and kick population constraints publicly available.
\end{abstract}

\maketitle

\begin{figure}[thb]
\includegraphics[width=0.3\textwidth]{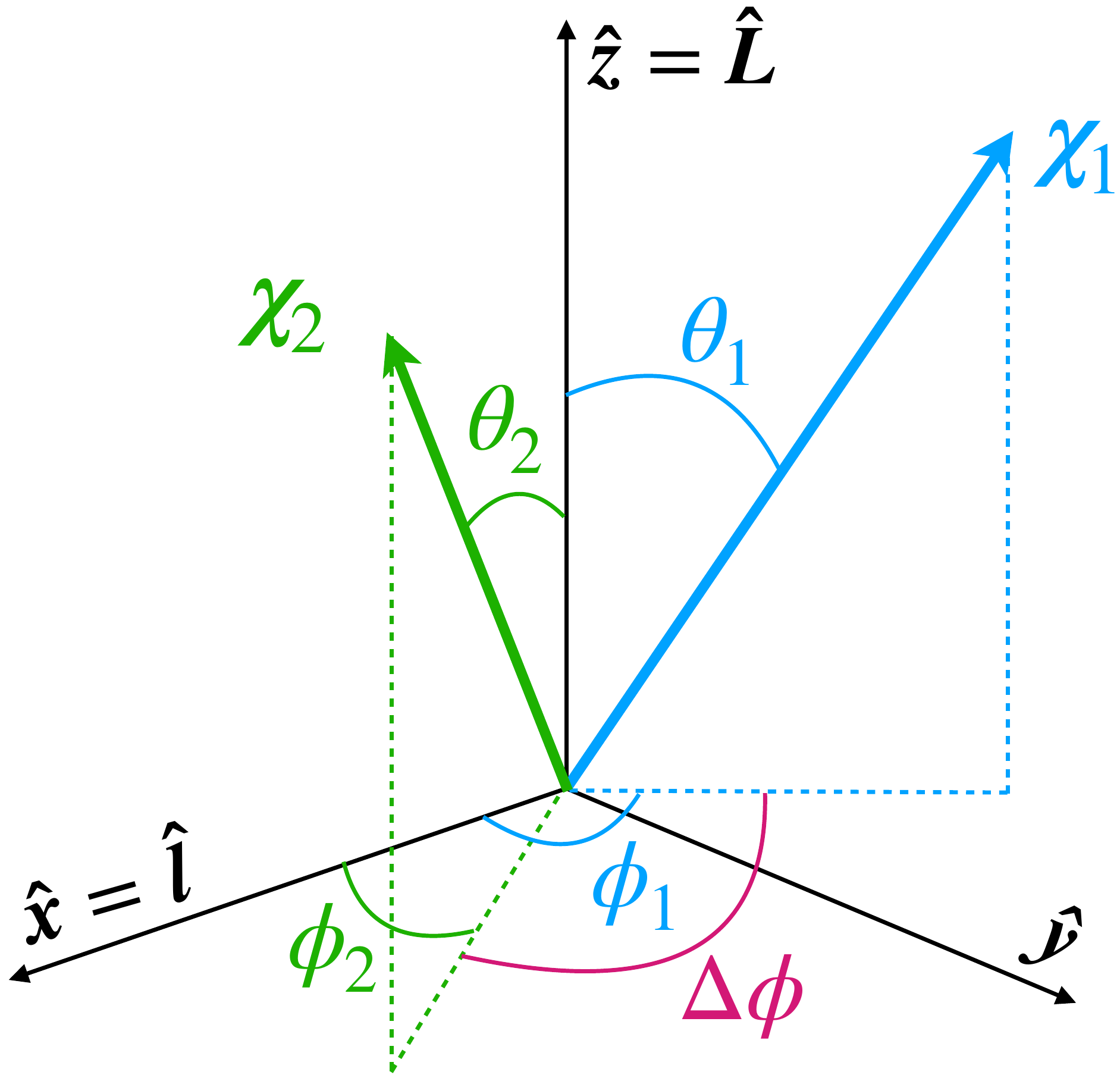}
\caption{
The BH spins are represented by 3-vectors $\bchi_1$ and $\bchi_2$, with index 1
(2) denoting the heavier (lighter) BH. We parameterize the spins by their
dimensionless magnitudes $\chi_1,\chi_2\leq1$, tilts $\theta_1,\theta_2$ w.r.t
the orbital angular momentum $\bL$~\cite{spinanglespaperfottnoteL}, and
orbital-plane spin angles $\phi_1,\phi_2$ w.r.t the line of separation $\bm{l}$
from the lighter to the heavier BH. Finally, $\delphi=\phi_1-\phi_2$.
}
\label{fig:spin_frames}
\end{figure}

\prlsec{Introduction.}
Binaries of spinning black holes (BHs) serve as a unique astrophysical
laboratory for a range of relativistic phenomena. For
example, if the BH spins $\bchi_1$ and $\bchi_2$ are aligned with the orbital
angular momentum $\bL$, the orientations of the orbital plane and the spins
remain fixed during the inspiral (cf.  Fig.~\ref{fig:spin_frames} for
definitions of the binary BH spin parameters).  However, if the spins are
tilted with respect to $\bL$, relativistic spin-orbit and spin-spin coupling
cause the orbital plane and the spins to precess~\cite{Apostolatos:1994pre,
Kidder:1995zr}.

While the tilt angles $\theta_1$ and $\theta_2$ control precession, the
orbital-plane spin angles $\phi_1$ and $\phi_2$ play a central role in binaries
undergoing spin-orbit resonances (SORs)~\cite{Schnittman:2004vq}. For these
binaries, the $\bchi_1$, $\bchi_2$ and $\bL$ vectors become locked into a
common resonant-plane such that $\delphi = \phi_1 - \phi_2$ is fixed at 0 or
$\pm \pi$ as the binary precesses.
Refs.~\cite{Kesden:2014sla, Gerosa:2015tea} pointed out that this locking is a
limiting case of librating states near $\delphi\sim$ 0 or $\pm \pi$.  For
simplicity, we will follow previous literature~\cite{Gerosa:2014kta,
Trifiro:2015zda, Afle:2018slw} and refer to the more general librating states
as SORs.
While evidence for precession has been found in the astrophysical binary BH
population~\cite{Abbott:2020gyp}, SORs have not yet been observed even though
they are expected in some astrophysical scenarios.  For example, stellar
binaries can cluster near these resonances if supernova natal kicks and stellar
tides are significant~\cite{Gerosa:2013laa, Gerosa:2018wbw}.

Another important relativistic effect that gets amplified for spinning binaries
is the gravitational recoil. Gravitational waves (GWs) can carry away linear
momentum from the binary, imparting a recoil or kick velocity to the merger
remnant~\cite{Bonnor:1961linmom, PhysRev.128.2471, Bekenstein:1973ApJ,
Fitchett:1983MNRAS}. These velocities can reach values up to $\sim5000$ km/s
for precessing binaries~\cite{Campanelli:2007cga, Gonzalez:2007hi,
Lousto:2011kp}, large enough to be ejected from any host
galaxy~\cite{Merritt:2004xa}. Kick measurements from GW
observations~\cite{Varma:2020nbm} can be used to constrain the formation of
heavy BHs via successive mergers~\cite{Gerosa:2021mno}. However, the kick
depends very sensitively on the orbital-plane spin
angles~\cite{Brugmann:2007zj}.

GW observations by LIGO~\cite{TheLIGOScientific:2014jea} and
Virgo~\cite{TheVirgo:2014hva} have enabled increasingly precise constraints on
the astrophysical distributions of BH spin magnitudes and
tilts~\cite{LIGOScientific:2018jsj, Abbott:2020gyp}, but the distributions of
the orbital-plane spin angles remain unconstrained. Constraining these
distributions would allow us to understand the prevalence of SORs and merger
kicks in nature. The biggest obstacle for this, however, is the difficulty in
measuring $\phi_1$, $\phi_2$ and $\delphi$ from individual GW events with
current detectors~\cite{Vitale:2014mka, Schmidt:2014iyl, Biscoveanu:2021nvg,
Gerosa:2014kta, Trifiro:2015zda, Afle:2018slw}.

However, in a companion
paper,~\citeauthor{Varma:2021csh}~\cite{Varma:2021csh}, we show
that this can be greatly improved by measuring the spins near the merger, in
particular, at a fixed \emph{dimensionless} reference time $\trefmHundredM$
before the peak of the GW amplitude, rather than the traditional choice of a
fixed GW frequency of $\frefTwentyHz$. Here $M=m_1+m_2$ is the total
(redshifted) mass of the binary with component masses $m_1 \geq m_2$, and we
set $G=c=1$.  Ref.~\cite{Varma:2021csh} shows that this improvement can
be attributed to the waveform being more sensitive to variations in the
orbital-plane spin angles near the merger.  In particular, measuring the spins
near the merger leads to improved constraints for $\phi_1$ and $\phi_2$ for
several events in the latest GWTC-2 catalog~\cite{Abbott:2020niy,
LIGOScientific:2018mvr, GWOSC_paper, GWOSC:GWTC, GWOSC:GWTC-2} released by the
LIGO-Virgo Collaboration. While the $\delphi$ measurements are not
significantly impacted, Ref.~\cite{Varma:2021csh} shows that this parameter
will also be better constrained with louder signals expected in the future.

In this \emph{Letter}, we use the spin constraints from
Ref.~\cite{Varma:2021csh} to perform the first measurement of the full spin
distribution in the astrophysical binary BH population.
In particular, we identify a preference for $\delphi \sim \pm \pi$, which can
be a signature of SORs. Next, given the spin population, we derive constraints
on the kick population.  Finally, we use the kick constraints to estimate the
fraction of merger remnants retained by globular and nuclear star clusters.

\prlsec{Methodology.}
The first step in our analysis is to estimate the binary BH parameters from
individual GW signals, which is done following Bayes'
theorem~\cite{Thrane:2019pe}:
\begin{gather}
    p(\Theta|d) \propto \Li(d|\Theta) \, \pi(\Theta),
\label{eq:Bayes_single_orig_prior}
\end{gather}
where $p(\Theta|d)$ is the \emph{posterior} probability distribution of the
binary parameters $\Theta$ given the observed data $d$, $\Li(d|\Theta)$ is the
\emph{likelihood} of the data given $\Theta$, and $\pi(\Theta)$ is the
\emph{prior} probability distribution for $\Theta$. The full set of binary
parameters $\Theta$ is 15 dimensional~\cite{Abbott:2020niy}, and includes the
masses and spins of the component BHs as well as extrinsic properties such as
the distance and sky location.

In this work, we use the posteriors samples from Ref.~\cite{Varma:2021csh},
obtained using the numerical relativity (NR) surrogate waveform model
\NRSur~\cite{Varma:2019csw}, with the spins measured at $\trefmHundredM$.
\NRSur accurately reproduces precessing NR simulations and is necessary to
reliably measure the orbital-plane spin angles~\cite{Varma:2021csh}.  GWTC-2
includes a total of 46 binary BH events.  However, because \NRSur only
encompasses ${\sim}20$ orbits before merger, it can only be applied to the
shorter signals with $M \gtrsim 60 \,M_{\odot}$~\cite{Varma:2019csw}. This
reduces our set of events to \numEv; these events are listed in Tab.~I of
Ref.~\cite{Varma:2021csh}.

\begin{figure*}[thb]
\includegraphics[width=0.5\textwidth]{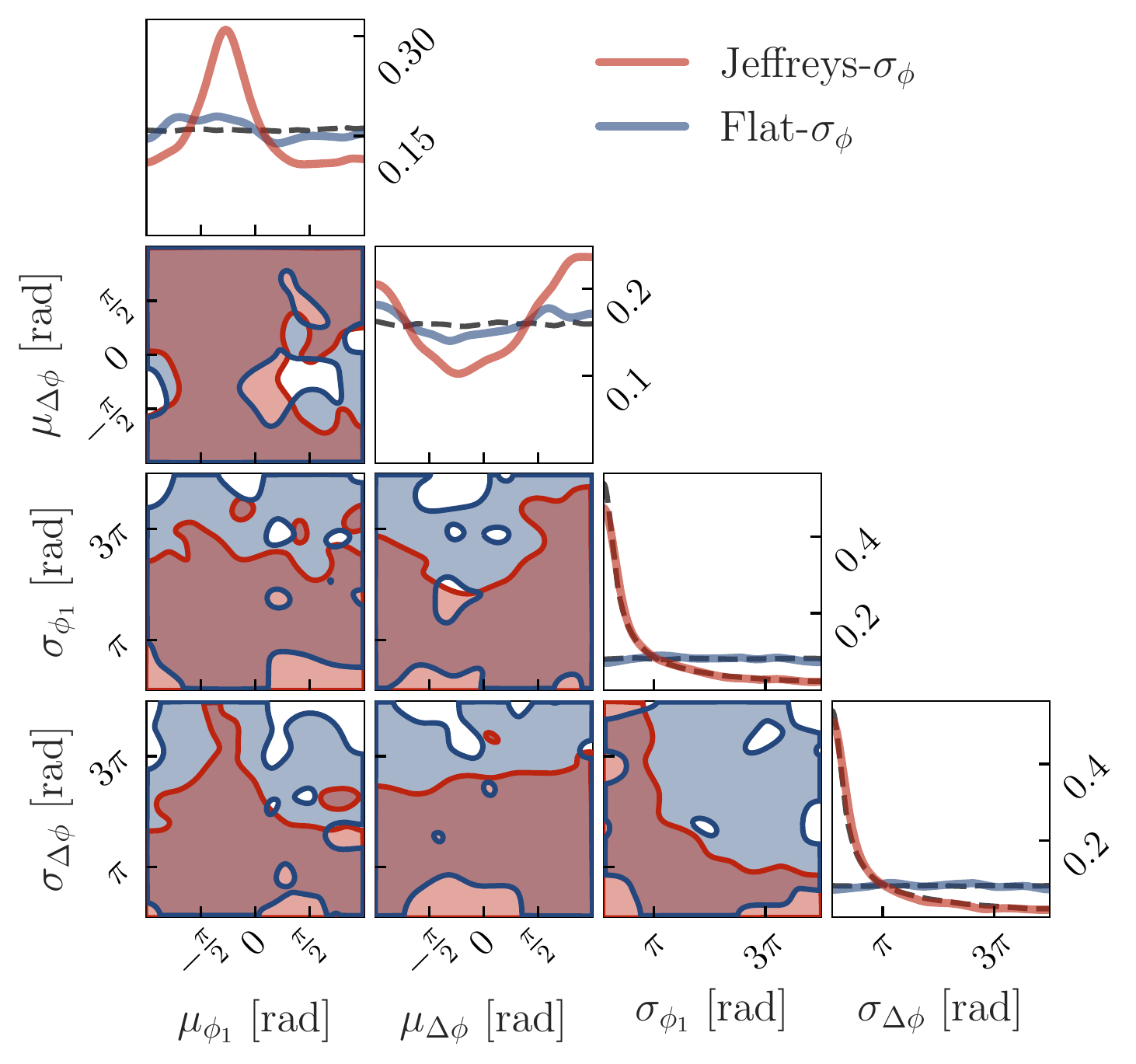}
\includegraphics[width=0.44\textwidth]{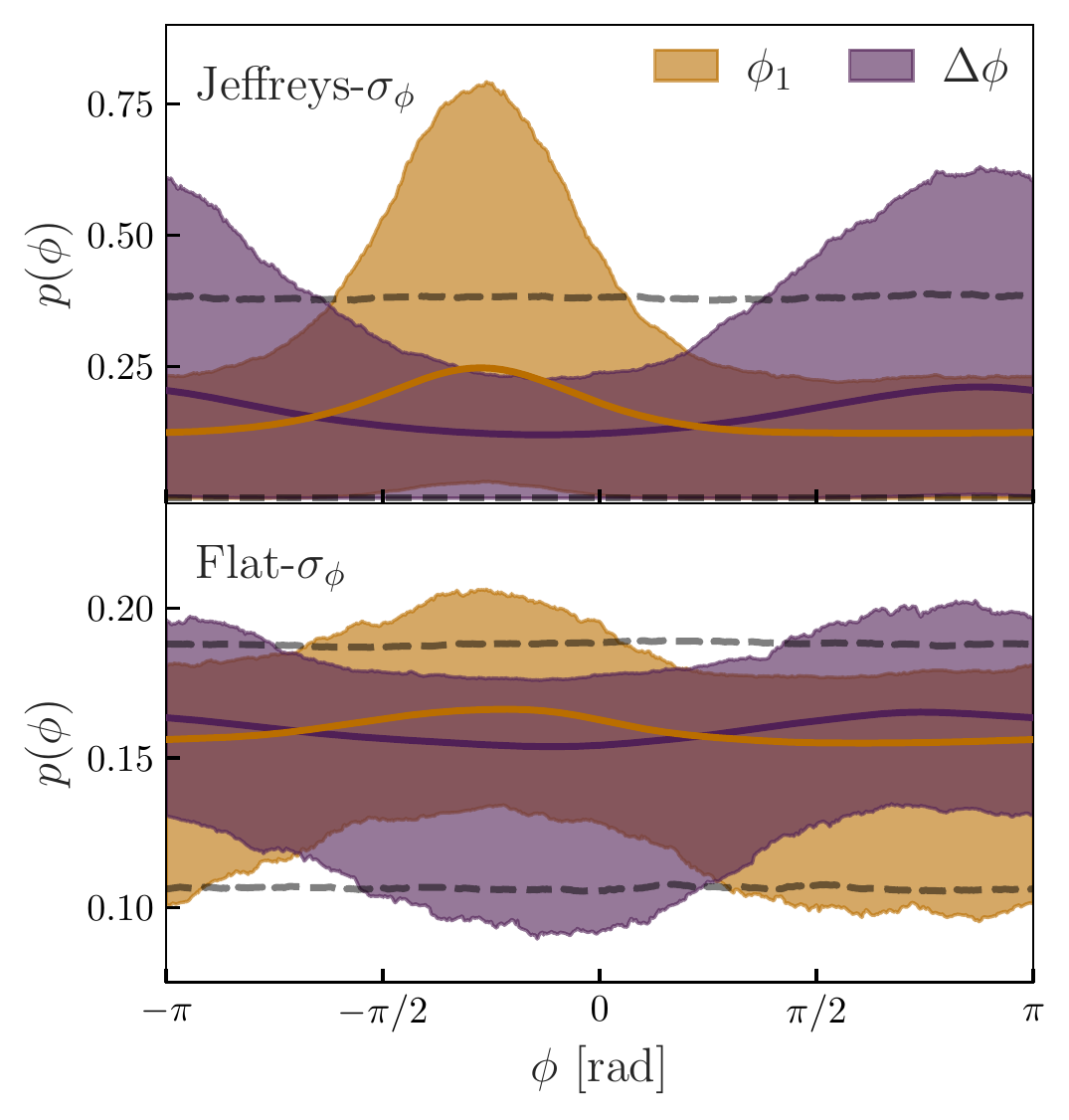}
\caption{
Constraints on the $\phi_1$ and $\delphi$ populations at
$\trefmHundredM$. \emph{Left:} Posteriors for the mean and width parameters.
The shaded regions show 90\% credible bounds on joint 2D posteriors. The
diagonal plots show 1D marginalized posteriors, with the priors shown as dashed
black lines. We consider two prior choices (\Jeffreys and \Flat) for the width
parameters. \emph{Right:} Constraints on the posterior population distributions
$p(\phi_1)$ and $p(\delphi)$ for the two prior choices. Shaded regions show
$90\%$ credible bounds, while the solid lines show the mean. The dashed grey
lines show the $90\%$ prior bounds.
}
\label{fig:spins_ppd}
\end{figure*}

Given the posterior samples $p(\Theta|d)$ for the individual events, we want to
measure the astrophysical distribution of the full spin degrees of freedom,
$\Spins = \{\chi_1, \chi_2, \theta_1, \theta_2, \phi_1, \delphi\}$, which is a
subset of $\Theta$. The remaining angle, $\phi_2$, is redundant given $\phi_1$
and $\delphi$; we choose to work with $\delphi$ as it is relevant for SORs. As
an intermediate step, we first reweight the posterior samples for each event to
account for known astrophysical constraints on the primary mass and mass ratio
($q=m_2/m_1$) populations~\cite{Abbott:2020gyp}. The details of the reweighting
procedure are given in the Supplement~\cite{spinanglespapersupplement}. Post
reweighting, Eq.~(\ref{eq:Bayes_single_orig_prior}) can be rewritten as:
\begin{gather}
    p(\Theta|d,\R) \propto \Li(d|\Theta) \, \pi(\Theta|\R),
\label{eq:Bayes_single_reweighted}
\end{gather}
where $\R$ indicates that these are the reweighted posteriors.
Using these reweighted posteriors in Eq.~(\ref{eq:hier_likelihood}) below
ensures that the implicit priors on the mass population are astrophysically
motivated~\cite{Vitale:2020aaz}.

To constrain the astrophysical distribution of $\Spins$, we begin by making the
assumption that the true value of $\Spins$ for each event is drawn from a
common underlying distribution $\pi(\Spins|\Lambda)$, which is conditional on a
set of \emph{hyperparameters} $\Lambda$. We then use hierarchical Bayesian
inference~\cite{Thrane:2019pe} to collectively analyze all \numEv events and
constrain $\Lambda$:
\begin{gather}
p(\Lambda|\{d_i\}) \propto \Li(\{d_i\}|\Lambda) \, \pi(\Lambda),
\label{eq:bayes_hier}
\end{gather}
where $p(\Lambda|\{d_i\})$ is the \emph{hyper-posterior} distribution for
$\Lambda$ given a set of observations $\{d_i\}$, $\Li(\{d_i\}|\Lambda)$ is the
\emph{hyper-likelihood} of this dataset given $\Lambda$, and $\pi(\Lambda)$ is
the \emph{hyper-prior} distribution for $\Lambda$. In our case, $\{d_i\}$ with
$i=1\ldots N$ represents the observed data for our set of $N=\numEv$ GW events.
The hyper-likelihood is obtained by coherently combining the data from from all
events~\cite{Thrane:2019pe}:
\begin{gather}
\Li(\{d_i\}|\Lambda) \propto  \prod_{i}^{N} \int d\Spins_i \,
    p(\Spins_i|d_i,\R) \,
    \frac{\pi(\Spins_i \,| \,\Lambda)}{\pi(\Spins_i|\R)} \,.
\label{eq:hier_likelihood}
\end{gather}

For the underlying distribution $\pi(\Spins|\Lambda)$, the spin magnitudes and
tilts are modeled following the ``Default spin'' model of
Ref.~\cite{Abbott:2020gyp}. The orbital-plane spin angles $\phi_1$ and
$\delphi$ are modeled as being drawn from independent von Mises
distributions~\cite{Mardia_Jupp_vonMises}. The von Mises distribution is an
approximation of a Gaussian distribution with periodic boundary conditions and
is parameterized by a mean and a standard deviation (or simply, width).

The explicit forms of $\pi(\Spins|\Lambda)$ and the hyper-prior $\pi(\Lambda)$
are given in the Supplement~\cite{spinanglespapersupplement}. In particular,
the priors on the mean ($\mu_{\phi_1}$ and $\mu_{\delphi}$) and width
($\sigma_{\phi_1}$ and $\sigma_{\delphi}$) hyperparameters for the $\phi_1$ and
$\delphi$ distributions are as follows. The prior for the mean parameters is
always uniform in $(-\pi, \pi)$. We consider two different prior choices for
the widths: (i) A Jeffreys prior~\cite{1946RSPSA.186..453J} that is log-uniform
in $\sigma_{\phi_1}$ and $\sigma_{\delphi}$ between $(0.3, 4\pi)$, henceforth
referred to as the \Jeffreys prior. (ii) A prior that is uniform in
$\sigma_{\phi_1}$ and $\sigma_{\delphi}$ between $(0.3, 4\pi)$, henceforth
referred to as the \Flat prior. The Jeffreys prior is an uninformative prior
choice often used for scale parameters~\cite{1946RSPSA.186..453J}. The flat
prior may be considered a control case to understand the impact of the prior.
The lower limit of $0.3$ rad on the width priors is arbitrary, but chosen to be
smaller than the sharpest features we expect to be resolvable with LIGO-Virgo
(which we estimate from the NR injections in Ref.~\cite{Varma:2021csh}). The
upper limit of $4\pi$ is chosen to be large enough to approximate a flat
distribution between $(-\pi, \pi)$.

We use the \texttt{Bilby}~\cite{Ashton:2018jfp} package with the
\texttt{dynesty}~\cite{Speagle:2019dynesty} sampler to draw posterior samples
for the hyperparameters $\Lambda$ from $p(\Lambda|\{d_i\})$.  Finally, the
posterior distribution for the $\Spins$ population, also referred to as the
\emph{posterior population distribution}, is obtained by averaging over
$\Lambda$~\cite{Thrane:2019pe}:
\begin{gather}
p(\Spins) = \int d\Lambda \, \pi(\Spins | \Lambda)
                 \, p(\Lambda|\{d_i\}).
\end{gather}
In practice, this is done by drawing samples from the hyper-posterior
$p(\Lambda|\{d_i\})$ and evaluating $\pi(\Spins | \Lambda)$ on an array of
$\Spins$ values for each $\Lambda$ sample.  This gives us an ensemble of
probability distributions on $\Spins$, which we use to compute the mean and
90\% credible widths. Finally, we note that we ignore selection effects for
the spin population, as they are not expected to be significant at current
sensitivity~\cite{spinanglespapersupplement}.

\prlsec{Spin population.}
Figure~\ref{fig:spins_ppd} shows our constraints on the $\phi_1$ and $\delphi$
populations. The left panel shows the posteriors for the mean and width
hyperparameters. For both \Jeffreys and \Flat prior choices, we find that the
1D marginalized posteriors for the widths $\sigma_{\phi_1}$ and
$\sigma_{\delphi}$ are dominated by the prior itself. However, the 1D
posteriors for the mean parameters show a preference for $\mu_{\phi_1} \sim
-\pi/4$ and $\mu_{\delphi} \sim \pm \pi$. This is reflected in the
corresponding constraints on the posterior population distributions,
$p(\phi_1)$ and $p(\delphi)$, shown in the right panel of
Fig.~\ref{fig:spins_ppd}. These represent our constraints on the astrophysical
distributions for $\phi_{1}$ and $\Delta\phi$; they are generated by evaluating
the von Mises model using draws from the joint posterior of the mean and width
parameters.

We interpret the population constraint in Fig.~\ref{fig:spins_ppd} as follows.
For the \Flat prior, examining the 2D posterior for $\mu_{\delphi} -
\sigma_{\delphi}$, we note that when $\sigma_{\delphi} \to 0$, only the region
around $\mu_{\delphi} \sim \pm \pi$ is allowed in the 90\% credible region.
This means that, if there is a sharp peak in the $\delphi$ population, it is
only allowed near $\sim \pm \pi$.  Similarly, examining the 2D posterior for
$\mu_{\phi_1} - \sigma_{\phi_1}$, we find that when $\sigma_{\phi_1} \to 0$,
there is a preference for $\mu_{\phi_1} \sim -\pi/4$.
The preferences in the 1D $\mu_{\phi_1}$/$\mu_{\delphi}$ posteriors and the
$p(\phi_1)$/$p(\delphi)$ distributions get amplified for the \Jeffreys prior,
as this prior already prefers small widths.
In short, the data disfavor peaks at regions other than $\phi_1 \sim -\pi/4$
and $\delphi \sim \pm \pi$, and this leads to $p(\phi_1)$ and $p(\delphi)$
peaks in these regions. However, the data are not informative enough to
constrain the widths of these peaks. We further note that both populations are
still consistent with a uniform distribution at the $90\%$ credible level.

It is important to recognize that the location of the $\phi_1$ peak in
Fig.~\ref{fig:spins_ppd} depends strongly on our choice of reference point.
This is because $\phi_1$ changes on the orbital timescale as it is defined with
respect to the line-of-separation (cf.~Fig.~\ref{fig:spin_frames}). On the
other hand, $\delphi$ only changes on the longer precession time scale, and we
find that repeating our analysis using spins measured at 20Hz leads to
consistent results for the $\delphi$
population~\cite{spinanglespapersupplement}. However, the biggest gain in
measuring the spins at $\trefmHundredM$ is in the $\phi_1$ population
constraint, as $\phi_1$ is significantly better measured
there~\cite{Varma:2021csh}. Constraining both $\phi_1$ and $\delphi$ is
necessary to constrain the kick population below.

For completeness, we include our constraints on the spin magnitude and tilt
populations, along with full model hyperparameter posteriors in the
Supplement~\cite{spinanglespapersupplement}.  Our constraints on the spin
magnitude and tilt populations are consistent with Ref.~\cite{Abbott:2020gyp},
and we do not find any obvious correlations between the orbital-plane spin
angles and the other spin parameters.  To gain further confidence in our
results, we also conduct some mock population
studies~\cite{spinanglespapersupplement}, which suggest that at least some
$\phi_1$ and $\delphi$ populations can be reliably recovered at current
detector sensitivity. Finally, by iteratively leaving one event out from the
dataset and repeating our analysis, we check that our results are not driven by
any single event.

One limitation of this work is the restriction to the \numEv signals with
$M\gtrsim60M_{\odot}$ so that we can use the \NRSur model. We also repeat our
analysis for all 46 binary BH events from GWTC-2, using the phenomenological
waveform model \PhenomT~\cite{Estelles:2021gvs} for the remaining 15 events.
Interestingly, we find that there is some information gain in the width
parameters in this case, with a preference for small widths. However, as noted
in Ref.~\cite{Varma:2021csh}, \PhenomT can have biases in recovering the
orbital-plane spin angles. Therefore, while we include these results in the
Supplement~\cite{spinanglespapersupplement} for completeness, we treat
Fig.~\ref{fig:spins_ppd} as our main result.

\prlsec{Kick population.}
Having constrained the full spin degrees of freedom for the binary BH
population, we can now derive constraints on the kick population.  We begin by
generating one realization of the $q$, $\bchi_1$ and $\bchi_2$ populations. For
$q$, we use the same model that was used in the initial posterior reweighting. \
For $\bchi_{1,2}$, we use the spin population constraints at $\trefmHundredM$.
We simply draw one hyperparameter sample from the posterior of the hierarchical
analysis and evaluate the $q$, $\bchi_1$ and $\bchi_2$ population models at
that point. Next, we draw a large number of samples for $q$, $\bchi_1$ and
$\bchi_2$ from this population realization and compute the corresponding kick
magnitudes using the \NRSurRemnant model~\cite{Varma:2019csw, Varma:2018aht}.
Repeating these steps over many draws of $q$, $\bchi_1$ and $\bchi_2$
populations, we generate an ensemble of kick population distributions $p(v_f)$.
For comparison, we also evaluate the prior $p(v_f)$ by repeating this procedure
using prior hyperparameter samples.

\begin{figure}[thb]
\includegraphics[width=0.47\textwidth]{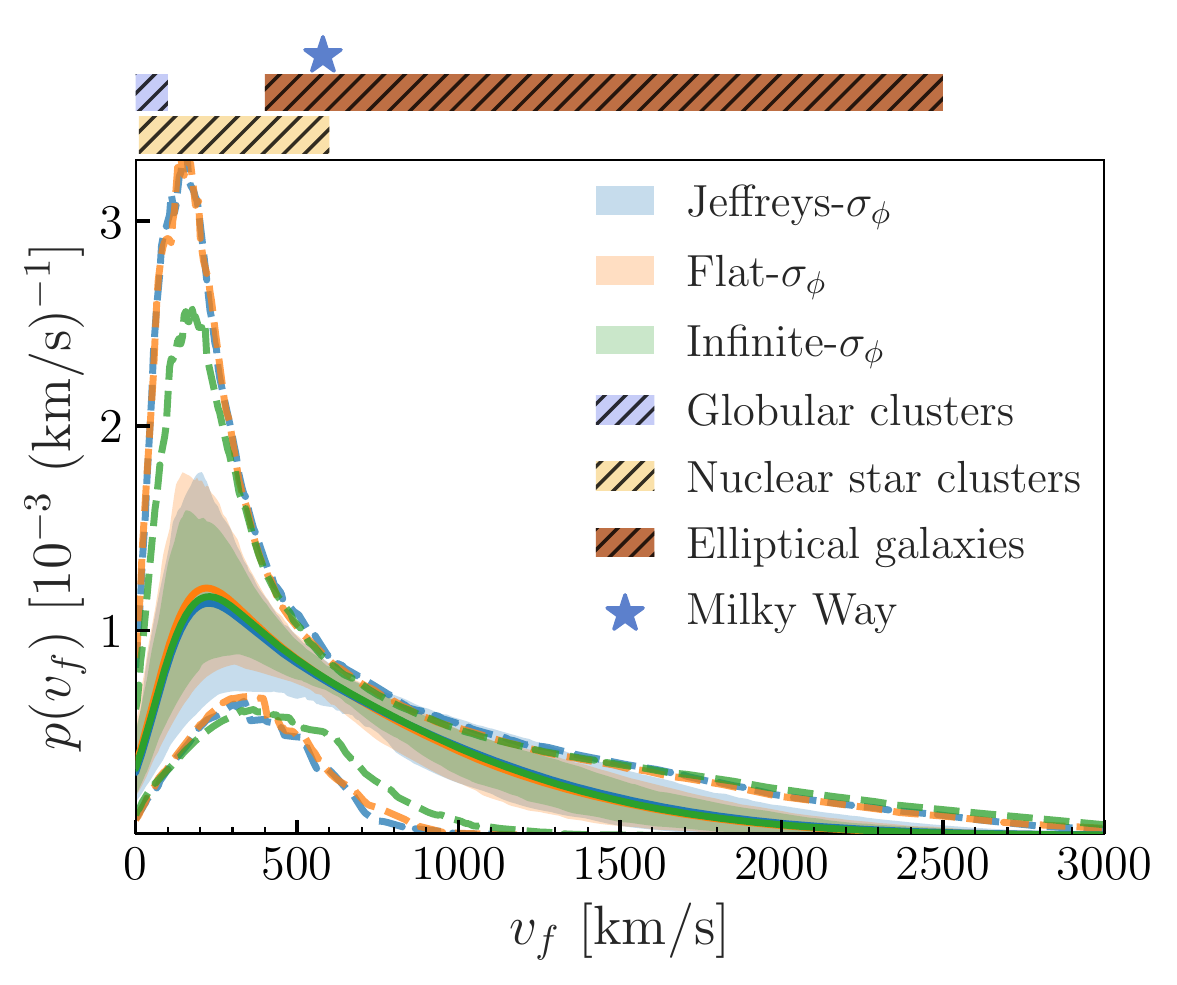}
\caption{
Constraints on the kick magnitude population for different prior choices for
$\sigma_{\phi_1}$ and $\sigma_{\delphi}$. Shaded regions show the central
$90\%$ credible bounds for the posterior, while the solid lines show the mean.
The dashed lines show the $90\%$ bounds for the prior. For comparison, we show
known ranges for the escape velocities for various types of host environments.
}
\label{fig:kick_ppd}
\end{figure}

Figure~\ref{fig:kick_ppd} shows the 90\% credible constraints on $p(v_f)$ for
the \Jeffreys and \Flat prior choices. In addition, we consider a prior choice
where the $\phi_1$ and $\delphi$ populations are restricted to be uniformly
distributed. We refer to this prior choice as \Infinite, as the other priors
reduce to this when $\sigma_{\phi_1}\!=\!\sigma_{\delphi}\!=\!\infty$. This
restricted prior was also used in Ref.~\cite{Doctor:2021qfn} to constrain the
kick population. The mass ratio, spin magnitude and tilt population models are
the same for all three choices. We compare our kick population constraints
against fiducial escape velocities for globular clusters~\cite{Gnedin:2002un,
Antonini:2016gqe}, nuclear star clusters~\cite{Antonini:2016gqe}, elliptical
galaxies~\cite{ Merritt:2004xa} and Milky Way-like
galaxies~\cite{Monari:2018esc}.

Comparing the prior and posterior ranges in Fig.~\ref{fig:kick_ppd}, we note
that there is significant information gain about the kick population, even
though individual events are largely uninformative about the
kick~\cite{Varma:2020nbm}.
The three prior choices lead to consistent kick populations in
Fig.~\ref{fig:kick_ppd}, with the \Infinite prior leading to the tightest
constraint. This is expected as the \Infinite prior is a special case of the
other two. This is also reflected in the more restrictive $p(v_f)$ prior in
Fig.~\ref{fig:kick_ppd} for \Infinite. It is somewhat surprising that the kick
population is not hugely influenced by the prior choices on $\phi_1$ and
$\delphi$ population, even though the kick is known to be very sensitive to
these parameters~\cite{Brugmann:2007zj}. This is explained by the fact that the
$\phi_1$ and $\delphi$ distributions in Fig.~\ref{fig:spins_ppd} are still
consistent with a uniform distribution at 90\% credibility. We expect this to
change with future observations.

\prlsec{Astrophysical implications.}
While the location of the $\phi_1$ peak in Fig.~\ref{fig:spins_ppd} is not
particularly important, the fact that there is a peak at all is indeed
interesting. This feature has not been predicted for any formation channel.
Therefore, our naive expectation is that this peak will get smoothed over as
more data are added. However, it will be interesting to see if there are
alternative explanations.

On the other hand, the preference for $\delphi \sim \pm \pi$ in
Fig.~\ref{fig:spins_ppd} is expected in some formation channels where
SORs~\cite{Schnittman:2004vq} are important. In particular, stellar binaries
with significant supernova natal kicks and efficient stellar tides can be
driven towards these resonances~\cite{Gerosa:2013laa, Gerosa:2018wbw}.  In the
standard scenario where the heavier star becomes the heavier BH, the $\delphi
\sim \pm \pi$ resonant mode is expected to be dominant. However, if mass
transfer between the two components is significant, a mass-ratio reversal
occurs and the $\delphi \sim 0$ mode becomes dominant~\cite{Gerosa:2013laa}.
Note that the predictions of Refs.~\cite{Gerosa:2013laa, Gerosa:2018wbw} are at
$\frefTwentyHz$, while our best constraints are at $\trefmHundredM$.  However,
as $\delphi$ only evolves on the precession timescale, we expect that a
preference for $\delphi \sim$ 0 or $\pm \pi$ at $\frefTwentyHz$ leads to a
similar preference at $\trefmHundredM$. It will be interesting to extend the
analysis of Refs.~\cite{Gerosa:2013laa, Gerosa:2018wbw} to $\trefmHundredM$,
for example, using the spin dynamics of \NRSur~\cite{Varma:2019csw}.

While our $\delphi$ population constraint can be interpreted as coming from
SORs, this does not yet constitute conclusive evidence for them---especially
not without a measurement of the width of the distribution to confirm this
feature. In addition, consider a binary for which the spin angular momenta in
the plane of the orbit perfectly cancel (which requires $\delphi=\pm \pi$).
This system will not undergo orbital precession and can approximately mimic a
spin-aligned system, as pointed out by Ref.~\cite{Biscoveanu:2021nvg}.  As a
result, a precessing waveform model can sometimes mistake a spin-aligned system
for one with $\delphi=\pm \pi$.  However, such a binary still undergoes spin
precession which can be used to break this degeneracy given sufficient
signal-to-noise ratio. A more detailed analysis may be necessary to account for
effects of such potential degeneracies on our results.

Finally, as an application of our $p(v_f)$ constraints in
Fig.~\ref{fig:kick_ppd}, we estimate the fraction of merger remnants that would
be retained by various host environments.  Assuming a maximum escape velocity
$\vesc\!=\!100$ km/s for globular clusters~\cite{Gnedin:2002un,
Antonini:2016gqe}, $6^{+5}_{-3}$ ($6^{+4}_{-3}$) \% of the remnants will be
retained for the \Jeffreys (\Flat) prior. For nuclear star clusters, assuming
$\vesc\!=\!600$ km/s~\cite{Antonini:2016gqe}, the retention fraction is
constrained to $54^{+19}_{-17}$ ($57^{+16}_{-15}$) \% for the \Jeffreys (\Flat)
prior. Averaging over the two prior choices, we estimate the retention fraction
to be $\sim \GCRange \, \%$ for globular clusters, and $\sim \NSCRange \, \%$
for nuclear star clusters. All constraints are quoted at 90\% credibility. Our
constraints on the retention fraction are consistent with those of
Refs.~\cite{Doctor:2021qfn, Mahapatra:2021hme}.

If the observed $\delphi \sim \pm \pi$ preference is confirmed to be due to
SORs, our findings have several important implications: (i) SOR measurements
can be used to place new astrophysical constraints on supernova natal kicks and
stellar tides~\cite{Gerosa:2013laa, Gerosa:2018wbw}. (ii) SORs can be used to
constrain the cosmic merger rate of isolated stellar binaries in galactic
fields, as well as measure what fraction of merging binaries form via that
channel. (iii) The $\delphi \sim \pm \pi$ resonance mode tends to enhance
merger kicks and suppress the spins of the remnant BHs~\cite{Berti:2012zp,
Kesden:2010ji}, both of which are important observables for constraining the
formation of heavy BHs via successive mergers~\cite{Gerosa:2021mno}.
Furthermore, independent of whether our findings can be attributed to SORs, our
$p(v_f)$ constraints already suggest that globular clusters are an unlikely
site for that formation channel.

\prlsec{Conclusion.}
We constrain the distribution of the orbital-plane spin orientations $\phi_1$
and $\delphi$ in the binary black hole population. We find that there is a
preference for $\delphi \sim \pm \pi$, which can be a signature of SORs. In
addition, we find a preference for $\phi_1 \sim -\pi/4$ in the population,
which has not been predicted for any astrophysical formation channel. However,
the strength of these preferences depends on our prior choices. Finally, we
constrain the distribution of recoil kicks in the population, and estimate the
fraction of merger remnants retained by globular and nuclear star clusters. We
make our population constraints publicly available at
Ref.~\cite{spinanglespaperdata}.

Observational evidence for SORs has far reaching implications for black hole
astrophysics. While our population constraints suggest the influence of SORs,
we are unable to constrain the widths of the $\phi_1$ and $\delphi$
distributions with the current dataset of events. Therefore, further
observations are necessary to confirm these trends.
With LIGO and Virgo approaching their design sensitivities~\cite{Aasi:2013wya},
our constraints are certain to improve in the near future.

\prlsec{Acknowledgments.}
We thank Davide Gerosa and Katerina Chatziioannou for useful discussions.
V.V.\ was supported by a Klarman Fellowship at Cornell. This project has
received funding from the European Union’s Horizon 2020 research and innovation
programme under the Marie Skłodowska-Curie grant agreement No.~896869.
S.B., M.I. and S.V. acknowledge support of the National Science Foundation
and the LIGO Laboratory.
S.B. is also supported by the NSF Graduate Research Fellowship under Grant No.
DGE-1122374.
M.I.\ is supported by NASA through the NASA Hubble Fellowship
grant No.\ HST-HF2-51410.001-A awarded by the Space Telescope
Science Institute, which is operated by the Association of Universities
for Research in Astronomy, Inc., for NASA, under contract NAS5-26555.
This research made use of data, software and/or web tools obtained from the
Gravitational Wave Open Science Center~\cite{GW_open_science_center}, a service
of the LIGO Laboratory, the LIGO Scientific Collaboration and the Virgo
Collaboration.
LIGO was constructed by the California Institute of Technology and
Massachusetts Institute of Technology with funding from the National Science
Foundation and operates under Cooperative Agreement No. PHY-1764464.
Computations were performed on the Alice cluster at ICTS; the Nemo cluster at
University of Wisconsin-Milwaukee, which is supported by NSF Grant PHY-1626190;
the Wheeler cluster at Caltech, which is supported by the Sherman Fairchild
Foundation and by Caltech; and the High Performance Cluster at Caltech.

\bibliography{References}

\clearpage
\section*{\large Supplemental materials}
\label{supp_mat}
\renewcommand{\spinframefignum}{\ref{fig:spin_frames}\xspace}
\renewcommand{\spinpopfignum}{\ref{fig:spins_ppd}\xspace}
\renewcommand{\kickpopfignum}{\ref{fig:kick_ppd}\xspace}
\renewcommand{\bayeseqnum}{\ref{eq:Bayes_single_orig_prior}\xspace}
\renewcommand{\reweightedbayeseqnum}{\ref{eq:Bayes_single_reweighted}\xspace}
\renewcommand{\hierbayeseqnum}{\ref{eq:bayes_hier}\xspace}
\renewcommand{\hierLieqnum}{\ref{eq:hier_likelihood}\xspace}
\setcounter{equation}{0}
\setcounter{figure}{0}
\setcounter{table}{0}
\renewcommand{\theequation}{S\arabic{equation}}
\renewcommand{\thefigure}{S\arabic{figure}}
\renewcommand{\thetable}{S\arabic{table}}

\section{Additional details on the hierarchical analysis}
\label{sec:app_hier}

In this section, we provide additional details on the reweighting procedure
applied to posterior samples for individual GW events, the explicit forms for
the spin population model and the hyper-prior, and potential selection effects
in the hierarchical analysis.

\subsection{Reweighting to an astrophysical mass, mass ratio, and redshift
prior}
\label{sec:app_reweighting}

We are interested in constraining the population for the full spin degrees of
freedom, $\Spins = \{\chi_1, \chi_2, \theta_1, \theta_2, \phi_1, \delphi\}$
(cf. Fig.~\spinframefignum). Recently, Ref.~\cite{Abbott:2020gyp}
analyzed the GWTC-2 catalog to place constraints on the populations of
component BH masses, spin magnitudes and tilts (but not $\phi_1$ and
$\delphi$). For simplicity, we only model the spin degrees of freedom in this
work, but incorporate the astrophysical mass and mass-ratio population
constraints from Ref.~\cite{Abbott:2020gyp}.

We denote $\Gamma=\{\mOneSrc, q\}$, where $\mOneSrc=m_1/(1+z)$ is the mass of
the heavier BH in the source frame, and $z$ is the source redshift. To account
for the astrophysical constraints on $\Gamma$, we apply the following weights
to the posterior samples for each individual event
\begin{align}
w(\Gamma_j) = \frac{p(\Gamma_{j} | \{d\}_{i\neq j})}{\pi(\Gamma_{j})},
\end{align}
where $j$ indicates the particular GW event, $\pi()$ is the same prior as in
Eq.~(\bayeseqnum), and $p(\Gamma_{j} | \{d\}_{i\neq j})$ denotes the posterior
population distribution (cf. Eq.~(46) of Ref.~\cite{Galaudage:2019jdx}) for the
$\Gamma$ population obtained using the data from all events other than $j$. We
use the public data release for the ``Power Law + Peak'' model from
Ref.~\cite{Abbott:2020gyp} for the $\Gamma$ population constraints.
$p(\Gamma_{j} | \{d\}_{i\neq j})$ is obtained from these results using the
``leave-one-out'' computation described in Ref.~\cite{Galaudage:2019jdx}. This
ensures that the event $j$ is not double-counted in the analysis.

We perform an additional reweighting to switch to a more astrophysically
motivated redshift prior. Note that, when obtaining the posterior samples in
Eq.~(\bayeseqnum), Ref.~\cite{Varma:2021csh} used a prior that is uniform in
comoving volume~\cite{Romero-Shaw:2020owr, Abbott:2020niy}:
\begin{align}
\pi(z) \propto \frac{dV_{c}}{dz},
\end{align}
where $dV_{c}/dz$ is the differential comoving volume. We now apply the weights
\begin{align}
w(z_j)  = (1+z_j)^{-1}.
\end{align}
to these posteriors, effectively switching to a prior that that is uniform in
comoving volume \emph{and} source frame time~\cite{Romero-Shaw:2020owr,
Abbott:2020niy}:
\begin{align}
\pi(z) \propto \frac{dV_{c}}{dz} \, (1+z)^{-1}.
\end{align}
The additional $(1+z)^{-1}$ factor accounts for cosmological time dilation.
This matches the redshift prior assumed for the ``Power Law + Peak'' model in
Ref.~\cite{Abbott:2020gyp}. We perform both reweighting steps simultaneously,
by applying the weights
\begin{align}
    w(\Theta_j) = w(\Gamma_j) \, w(z_j),
\end{align}
to the posterior samples $\Theta_j$ for each event $j$ in our dataset. Post
reweighting, Eq.~(\bayeseqnum) can be rewritten as Eq.~(\reweightedbayeseqnum).

Note that Ref.~\cite{Abbott:2020gyp} models the mass degrees of freedom
simultaneously with the spin degrees of freedom. Using the
mass-population-reweighted posteriors is equivalent to
Ref.~\cite{Abbott:2020gyp}, except that this does not account for any possible
correlations between the mass and spin hyperparameters. However,
Ref.~\cite{Abbott:2020gyp} found that these correlations are not significant.

\begin{table}[tbh]
    \centering
\begin{tabular}{| c  l |}
    \hline
    Parameter & Prior\\
    \hline
    $\mu_{\chi}$ & U(0, 1) \\
    $\sigma^2_{\chi}$  & U(0, 0.25) \\
    $\xi_{\theta}$ & U(0,1) \\
    $\sigma_{\theta}$  & U(0.01,4) \\
    $\mu_{\phi_1}$ & U($-\pi$, $\pi$) \\
    $\mu_{\delphi}$ & U($-\pi$, $\pi$) \\
    $\sigma_{\phi_1}$ & See Tab.~\ref{tab:hyperpriors_phi}.  \\
    $\sigma_{\delphi}$ & See Tab.~\ref{tab:hyperpriors_phi}. \\
    \hline
\end{tabular}
\caption{Priors on hyperparameters for our spin population model. In addition,
following Ref.~\cite{Abbott:2020gyp}, we exclude $\mu_{\chi}, \sigma^2_{\chi}$
values where the Beta distribution becomes singular. Here, U($a$, $b$)
indicates a uniform distribution on the interval ($a$, $b$).
}
\label{tab:hyperpriors}
\end{table}

\begin{table}[tbh]
    \centering
    \begin{tabular}{| l c |}
    \hline
    Name & Prior on $\sigma_{\phi_1}$ and $\sigma_{\delphi}$\\
    \hline
    \Jeffreys & LU(0.3, $4\pi$)  \\
    \Flat & U(0.3, $4\pi$)  \\
    \Infinite & $\delta(\infty)$  \\
    \hline
\end{tabular}
\caption{The prior choices we consider for $\sigma_{\phi_1}$ and
$\sigma_{\delphi}$. LU($a$, $b$) indicates a log-uniform distribution on
the interval ($a$, $b$), while $\delta(a)$ indicates a Dirac delta distribution
where the parameter is fixed at $a$. Note that the \Infinite prior restricts
the $\phi_1$ and $\delphi$ populations to be uniform.}
\label{tab:hyperpriors_phi}
\end{table}

\begin{figure*}[thb]
\includegraphics[width=0.93\textwidth]{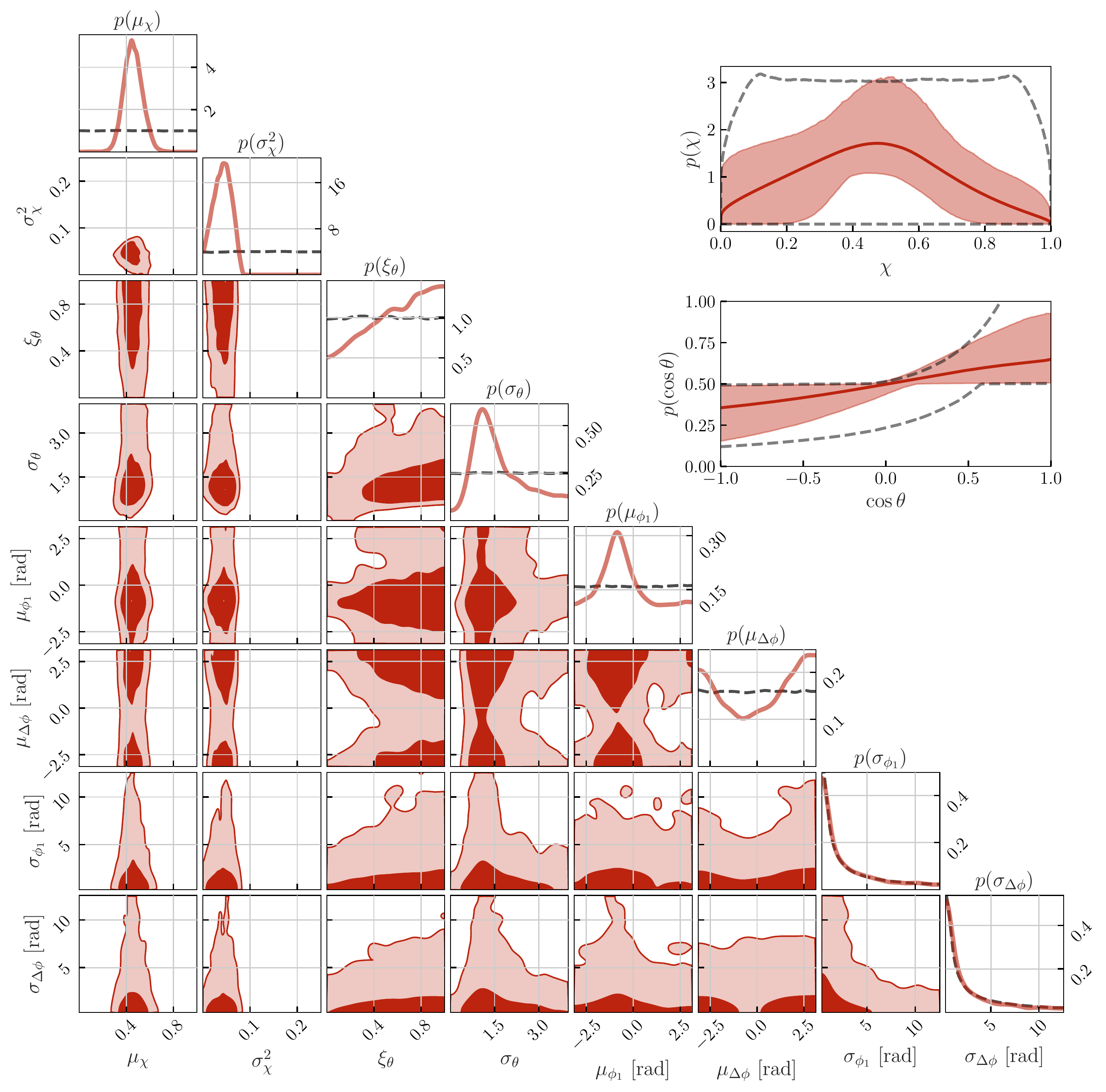}
\caption{
Posterior distribution for hyperparameters for our spin model with a Jeffreys
prior for $\sigma_{\phi_1}$ and $\sigma_{\delphi}$. The shaded regions in the
lower-triangle subplots represent $50\%$ and $90\%$ credible bounds on joint 2D
posteriors. The diagonal subplots show the 1D marginalized posteriors and
priors (black dashed lines). In the top-right, we show constraints on the spin
magnitude and tilt populations. Shaded regions show the central $90\%$ credible
bounds, while the solid lines show the mean. The dashed grey lines show the
$90\%$ prior bounds. The corresponding population constraints on $\phi_1$ and
$\delphi$ are shown in the top half of the right-panel of
Fig.~\spinpopfignum.
}
\label{fig:corner_Jeffreys}
\end{figure*}

\begin{figure*}[thb]
\includegraphics[width=0.93\textwidth]{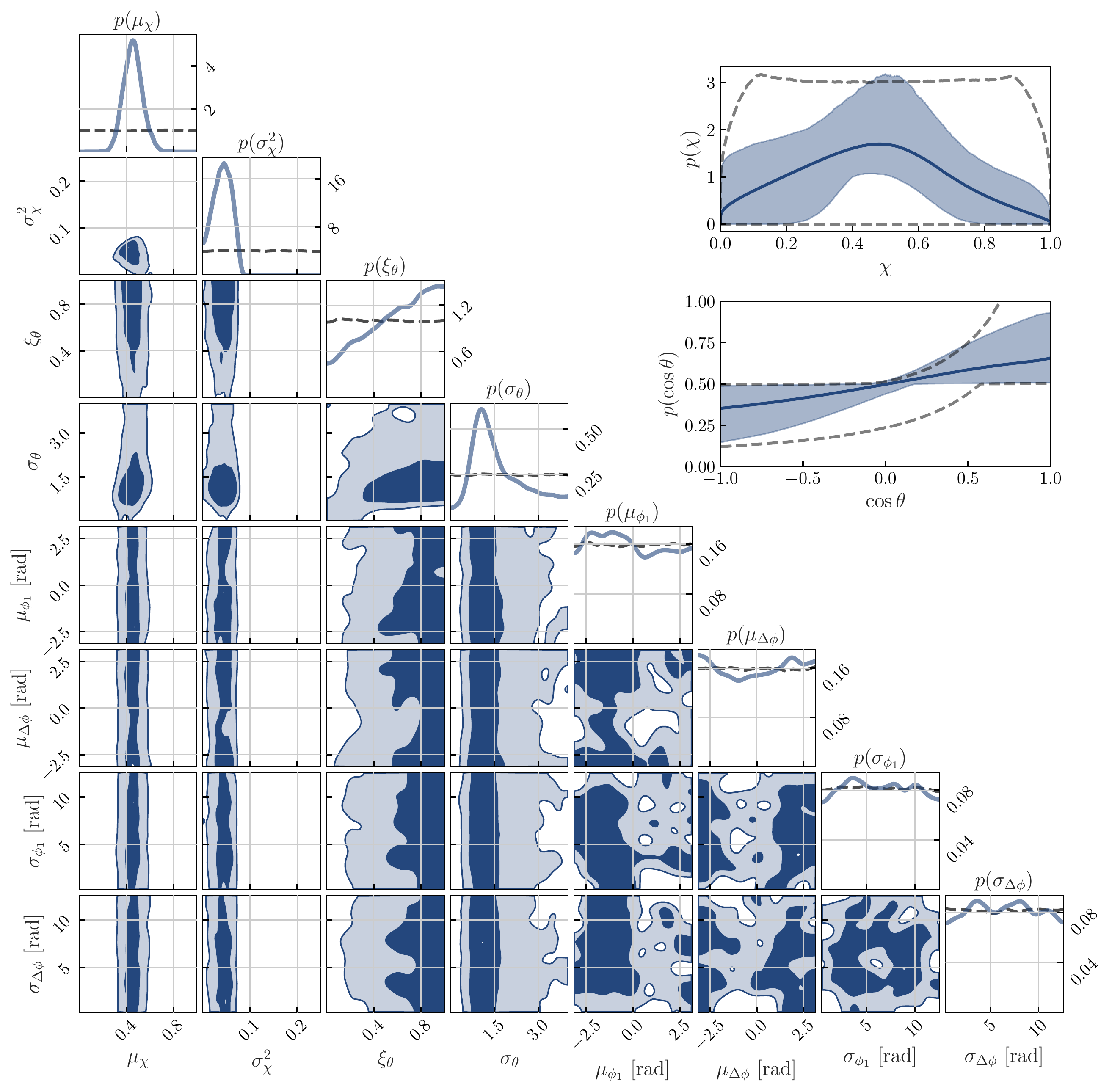}
\caption{
Same as Fig.~\ref{fig:corner_Jeffreys}, but using a Flat prior for
$\sigma_{\phi_1}$ and $\sigma_{\delphi}$. The corresponding $\phi_1$ and
$\delphi$ population constraints are shown in the bottom half of the
right-panel of Fig.~\spinpopfignum.
}
\label{fig:corner_Flat}
\end{figure*}

\subsection{Population model}
We use the following joint distribution for the underlying spin distribution
$\pi(\Spins|\Lambda)$ in Eq.~(\hierLieqnum):
\begin{align}
\pi(\Spins|\Lambda)
    & = p(\chi_{1,2} \, | \,\mu_{\chi},\sigma^2_{\chi}) ~
        p(\theta_{1,2} \, | \, \xi_{\theta}, \sigma_{\theta}) \nonumber \\
    & ~~~~ p(\phi_1 \, | \, \mu_{\phi_1}, \sigma_{\phi_1}) ~
        p(\delphi \, | \, \mu_{\delphi}, \sigma_{\delphi}).
\end{align}
Here, $p(\chi_{1,2} \, | \,\mu_{\chi},\sigma^2_{\chi})$ is a Beta distribution
in the spin magnitudes, parameterized by its mean $\mu_{\chi}$ and variance
$\sigma^2_{\chi}$~\cite{Wysocki:2018mpo}, and $p(\theta_{1,2} \, | \,
\xi_{\theta}, \sigma_{\theta})$ is an isotropic tilt distribution with a
Gaussian peak component, parameterized by the standard deviation
$\sigma_{\theta}$ of the Gaussian and the mixing fraction $\xi_{\theta}$ coming
from the Gaussian component~\cite{Talbot:2017yur}. Note that we assume the
distributions for the two component BHs are the same for the spin magnitude and
tilt. This model for the spin magnitudes and tilts is the same as ``Default
spin'' model described in App.D.1 of Ref.~\cite{Abbott:2020gyp}. On top of this
model, we include $p(\phi_1 \, | \, \mu_{\phi_1}, \sigma_{\phi_1})$ and
$p(\delphi \, | \, \mu_{\delphi}, \sigma_{\delphi})$ as independent von Mises
distributions parameterized by their corresponding mean and standard
deviations. Our choices for the hyper-prior $\pi(\Lambda)$ (cf.
Eq.~(\hierbayeseqnum)) imposed on the hyperparameters are described in
Tab.~\ref{tab:hyperpriors}.

The von Mises distribution~\cite{Mardia_Jupp_vonMises} is defined as
\begin{gather}
p(\phi | \mu, \kappa)
    = \frac{\exp{\left(\kappa \, \cos{\left(\phi - \mu\right)} \right)}}
    {2 \pi \, I_0(\kappa)},
\end{gather}
where $\mu$ is the mean, $\kappa$ is a shape parameter, and $I_0$ is the
modified Bessel function of order 0. The von Mises distribution is a close
approximation of a Gaussian with periodic boundary conditions at $\phi=\pm
\pi$, making it an appropriate choice for phase parameters like $\phi_1$ and
$\delphi$. The variance of the von Mises distribution can be approximated as
$1/\kappa$; therefore we define the standard deviation to be $\sigma \equiv
1/\sqrt{\kappa}$.

\subsection{Selection effects}
Ref.~\cite{Abbott:2020gyp} also distinguishes between between the
\emph{astrophysical} distribution of a parameter— the distribution as it is in
nature—and the \emph{observed} distribution of a parameter—the distribution as
it appears among detected events due to selection effects, because of which
binaries with certain parameters may be easier to detect than others. These
effects can be accounted for by modifying the hyper-likelihood in
Eq.~(\hierLieqnum) as done in Eq.~(1) of Ref.~\cite{Abbott:2020gyp}. While
Ref.~\cite{Abbott:2020gyp} includes selection effects for their mass population
models, they are ignored for the ``Default spin'' model as they are not
expected to be significant at current detector sensitivity.

In our case, the mass-reweighted posteriors (cf.
Sec.~\ref{sec:app_reweighting}) already account for selection effects for the
mass population. As we simply extend the spin population model of
Ref.~\cite{Abbott:2020gyp} with the $\phi_1$ and $\delphi$ models, we also
ignore selection effects for our spin population model. As noted in
Ref.~\cite{Abbott:2020gyp}, it will be important to include spin selection
effects as detectors sensitivity improves. However, at current sensitivity,
assuming spin selection effects are not significant (as done in
Ref.~\cite{Abbott:2020gyp}), our results can be treated as constraints on the
astrophysical distribution rather than the observed distribution.

Additional selection effects can arise from the \NRSur requirement of $M
\gtrsim60M_{\odot}$, which restricts us to \numEv of the available 46 events in
GWTC-2. However, assuming the correlations between the mass and spin population
distributions are not significant at current sensitivity~\cite{Abbott:2020gyp},
imposing an additional selection criterion on the masses should not affect the
inferred spin distribution, as the \numEv events included in our analysis would
be a fair representation of the underlying spin distribution. Note that while
Ref.~\cite{Abbott:2020gyp} found that such correlations are not significant,
this did not include the orbital-plane spin angles. Therefore, a full
resolution of this would require extending \NRSur to longer inspirals; see
Ref.~\cite{Varma:2018mmi} for work in this direction. Finally, \NRSur is also
restricted to mass ratios $q \geq 1/6$~\cite{Varma:2019csw}. However, this
restriction does not exclude any additional events, as the only GWTC-2 events
with significant support at $q\lesssim 1/6$ also have a total mass $< 60
M_{\odot}$~\cite{Abbott:2020niy}.

\section{Additional investigations}

\subsection{Full spin population}
In Fig.~\spinpopfignum, we only show the population constraints on $\phi_1$
and $\delphi$. For completeness, we now show the hyperparameter posteriors and
population constraints on the spin magnitudes and tilts in
Fig.~\ref{fig:corner_Jeffreys} (for the \Jeffreys prior) and
Fig.~\ref{fig:corner_Flat} (for the \Flat prior). Our population constraints on
the spin magnitudes and tilts are consistent with Ref.~\cite{Abbott:2020gyp},
but somewhat broader as we only use \numEv of the available 46 binary BH events
in GWTC-2. We do not find any clear correlations between the orbital-plane
spin angles and the other spin parameters.

The spin magnitude and tilt populations of Ref.~\cite{Abbott:2020gyp} are
constrained at $\frefTwentyHz$ while our constraints are at $\trefmHundredM$.
However, we expect these populations to be similar, as spin tilt measurements
at current detector sensitivity are not strongly dependent on the reference
point~\cite{Varma:2021csh}. While the BH spin magnitudes can evolve during the
inspiral due to in-falling angular momentum carried by GWs, this is a very
small effect (4PN higher than leading angular-momentum loss~\cite{Alvi:2001mx,
Poisson:2004cw}), and is thus safely ignored by current waveform models
including \NRSur. The orbital-plane spin angle measurements, on ther other
hand, do strongly depend on the reference point~\cite{Varma:2021csh}, which
leads to noticeable differences in the population constraints as discussed in
Sec.~\ref{sec:app_pop_20Hz}.

\subsection{Population constraints at $\frefTwentyHz$}
\label{sec:app_pop_20Hz}
\begin{figure}[thb]
\includegraphics[width=0.45\textwidth]{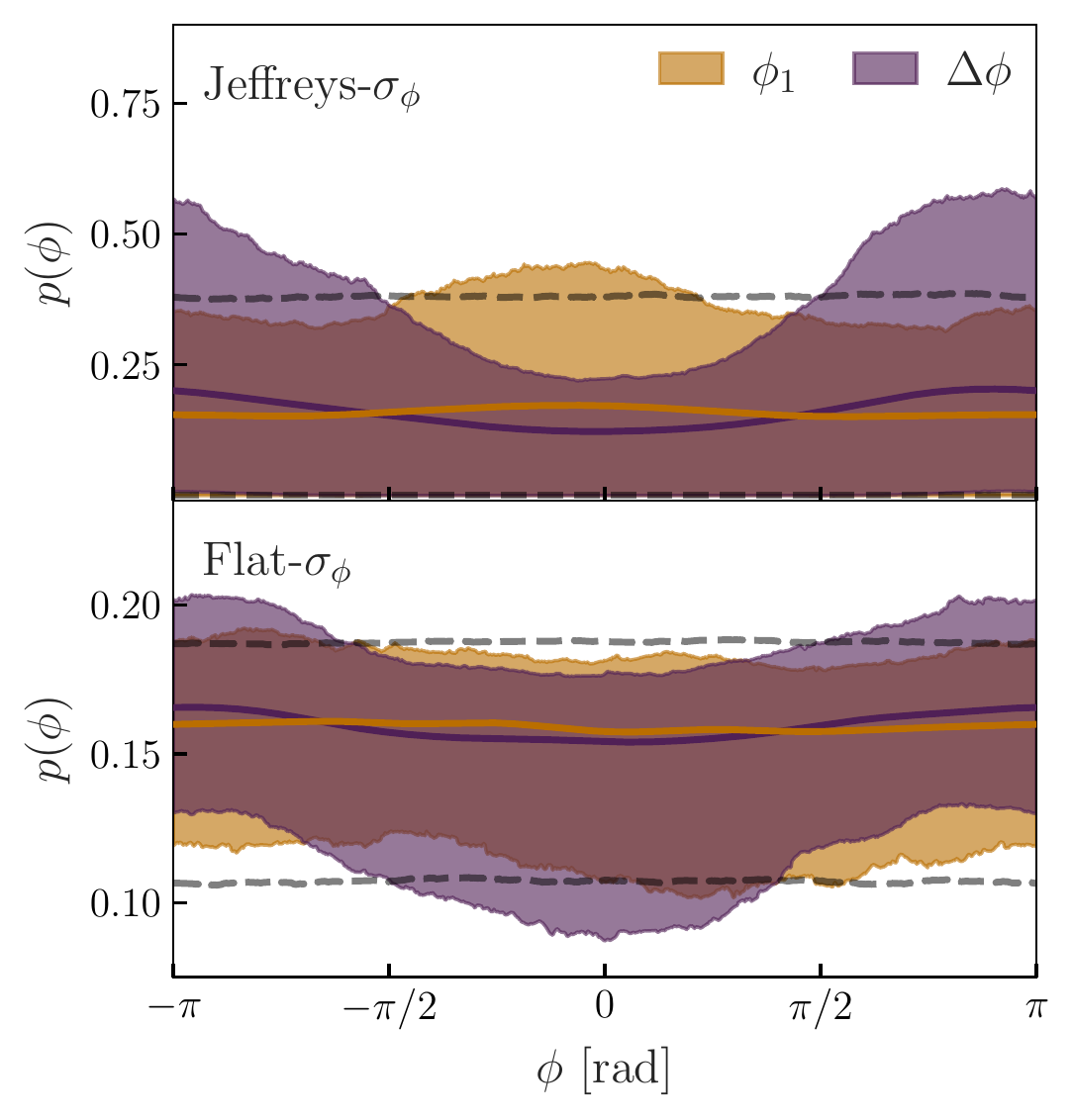}
\caption{
Same as Fig.~\spinpopfignum, but for spins measured at $\frefTwentyHz$.
While the $\delphi$ constraints are consistent between $\frefTwentyHz$ and
$\trefmHundredM$, $\phi_1$ is much better constrained at $\trefmHundredM$.
}
\label{fig:spins_ppd_20Hz}
\end{figure}

The results in Fig.~\spinpopfignum are obtained using \NRSur spin posteriors at
$\trefmHundredM$~\cite{Varma:2021csh}. Ref.~\cite{Varma:2021csh} also generated
\NRSur spin posteriors at $\frefTwentyHz$. In this section, we repeat our
hierarchical analysis using these spin posteriors for comparison.
Figure~\ref{fig:spins_ppd_20Hz} shows constraints on the $\phi_1$ and $\delphi$
populations when spins are measured at $\frefTwentyHz$. Because the $\delphi$
measurements for individual events do not change significantly between
$\frefTwentyHz$ and $\trefmHundredM$ (cf.~Fig.4 of Ref.~\cite{Varma:2021csh}),
the $\delphi$ populations are also consistent between
Fig.~\ref{fig:spins_ppd_20Hz} and Fig.~\spinpopfignum.  By contrast, as there
is significant improvement in $\phi_1$ measurements for individual events at
$\trefmHundredM$ (cf.~Fig.3 of Ref.~\cite{Varma:2021csh}), the $\phi_1$
population is much better constrained in Fig.~\spinpopfignum compared to
Fig.~\ref{fig:spins_ppd_20Hz}. It is important to note that this does not imply
that the astrophysical $\phi_1$ distribution is flatter at $\frefTwentyHz$
compared to $\trefmHundredM$. Instead, this is because the $\phi_1$
measurements at $\frefTwentyHz$ are very poor. In fact, even the mild peak near
$\phi_1 \sim 0$ for the \Jeffreys prior in Fig.~\ref{fig:spins_ppd_20Hz} is
driven entirely by GW190521~\cite{Abbott:2020tfl}. As discussed in
Ref.~\cite{Varma:2021csh}, GW190521 is the only event with a good measurement
of $\phi_1$ at $\frefTwentyHz$ as this binary happens to merge near 20 Hz.

\begin{figure*}[thb]
\includegraphics[width=0.45\textwidth]{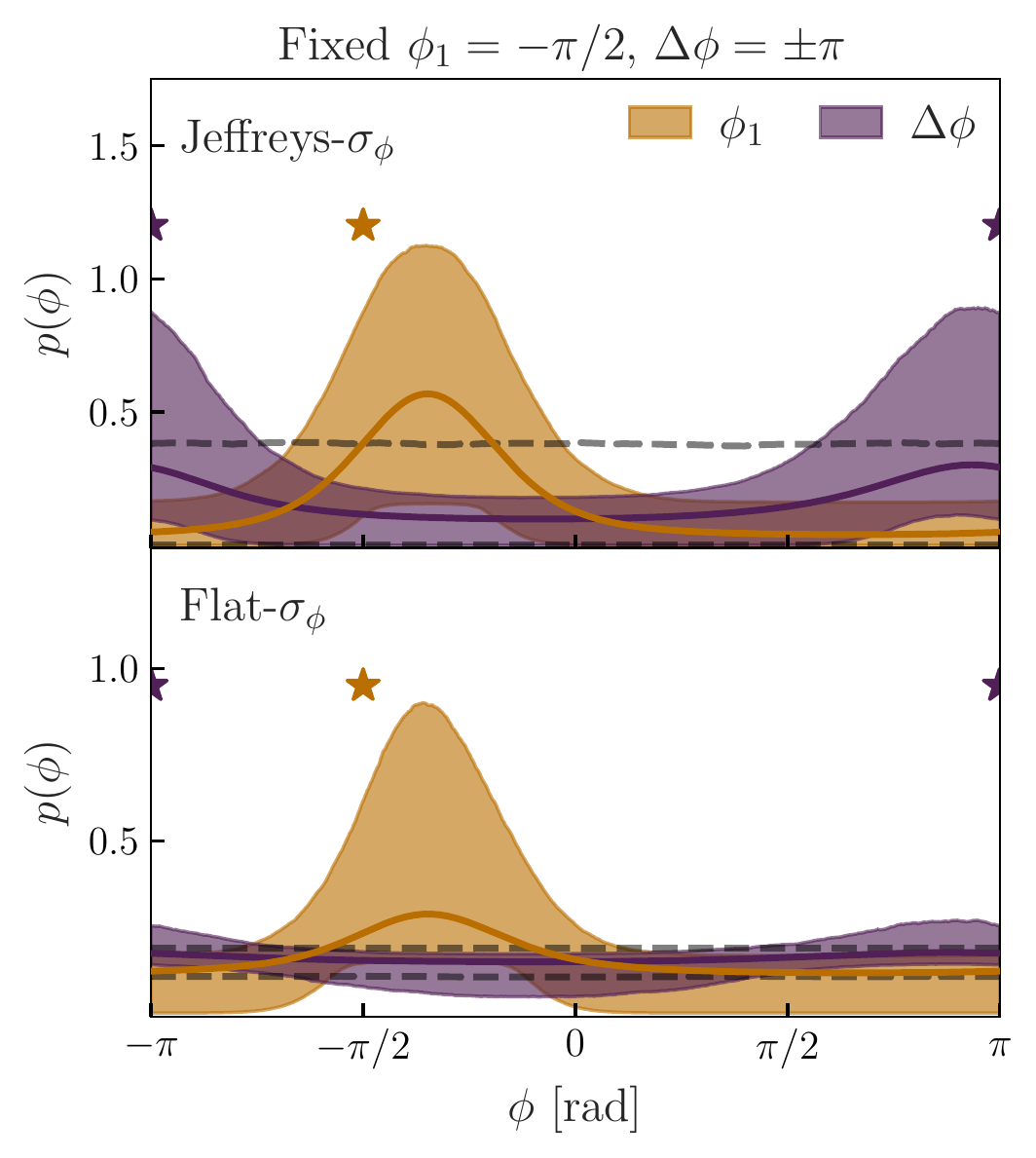}
\includegraphics[width=0.45\textwidth]{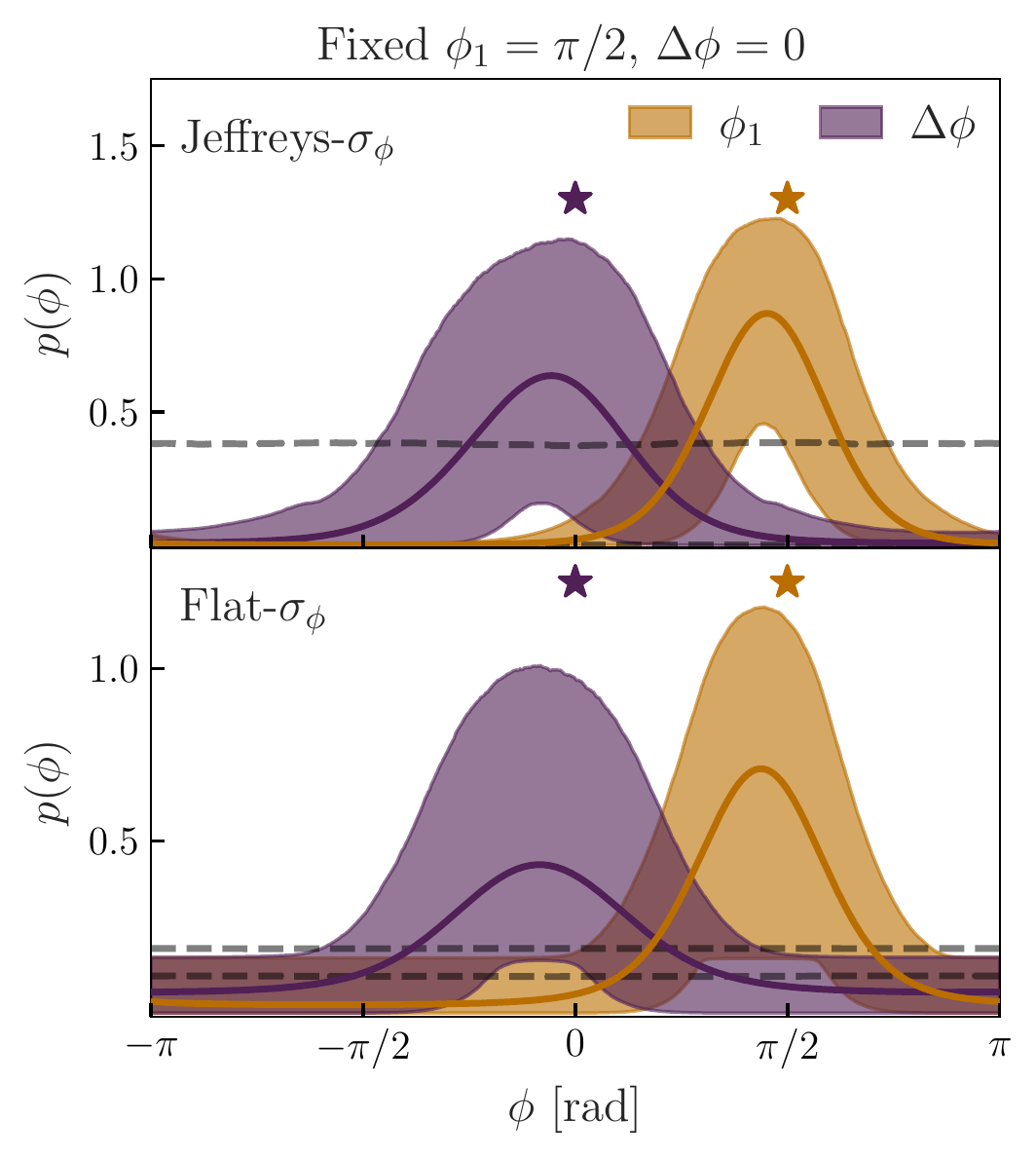}
\caption{
Same as Fig.~\spinpopfignum, but for the two mock populations (with \numEv
events) described in Sec.~\ref{sec:app_mock_pop}. The left panels correspond to
injections with fixed $\phi_1=-\pi/2$ and $\delphi=\pm \pi$ at
$\trefmHundredM$, while the right panels have fixed $\phi_1=\pi/2$ and
$\delphi=0$. The injected values are indicated by star markers. For both mock
populations, the true values are reasonably well recovered, but the $\delphi=0$
population is better constrained than the $\delphi = \pm \pi$ one, especially
for the \Flat prior.
}
\label{fig:spins_ppd_inj}
\end{figure*}

\begin{figure*}[thb]
\includegraphics[width=0.48\textwidth]{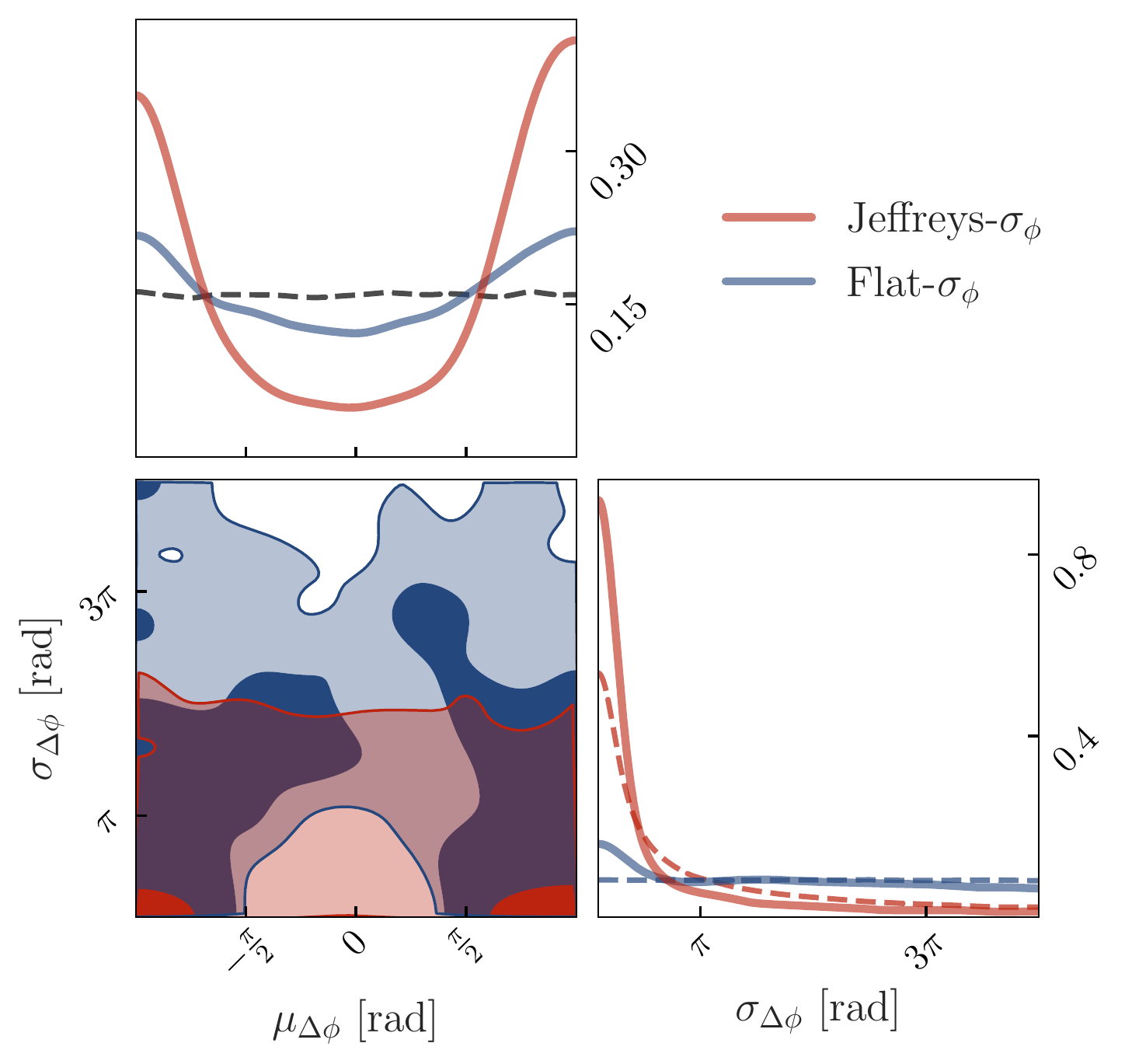} ~~~~~
\includegraphics[width=0.44\textwidth]{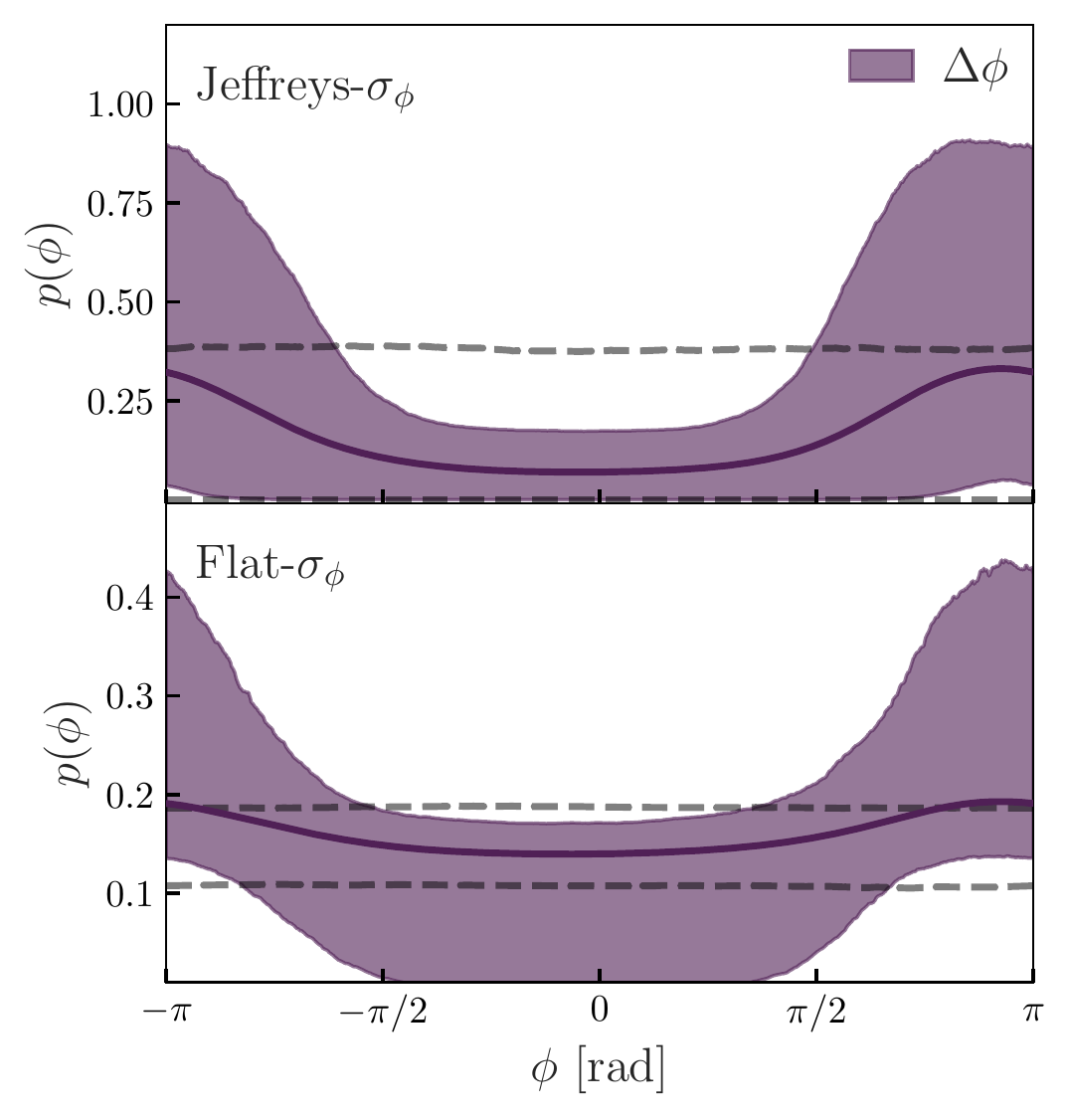}
\caption{
Constraints on the $\delphi$ population at $\frefTwentyHz$ using all 46 binary
BH events from GWTC-2. We use the \NRSur model for the \numEv events with $M
\gtrsim60M_{\odot}$, and the \PhenomT model for the remaining 15 events.
\emph{Left:} Posteriors for the mean and width parameters for $\delphi$.  The
shaded regions show 50\% (dark shade) and 90\% (light shade) credible bounds on
the joint 2D posterior. The top and right subplots show 1D marginalized
posteriors as solid lines. The prior on $\mu_{\delphi}$ is shown as a dashed
black line, while the two prior choices for $\sigma_{\delphi}$ are shown as
colored dashed lines. Unlike Fig.~\spinpopfignum, we now see some information
gain in the 1D posteriors for the width parameters, with a preference towards
small widths. \emph{Right:} Corresponding constraints on the $p(\delphi)$
population distribution. The peak at $\delphi \sim \pm \pi$ is amplified
compared to Fig.~\spinpopfignum.
}
\label{fig:spins_ppd_allEv}
\end{figure*}

Using the spins measured at $\frefTwentyHz$ to generate the kick population
results in distributions very similar to Fig.~\kickpopfignum of the main text.
Once again, this is explained by the fact that the $\phi_1$ and $\delphi$
distributions at both $\trefmHundredM$ and $\frefTwentyHz$ are still consistent
with a uniform distribution at 90\% credibility, at current sensitivity.

To project measurements of the kick population with improved detector
sensitivity, it is important to first appreciate that kick measurements for
individual events do not depend on the reference point when using the method in
Ref.~\cite{Varma:2020nbm}. The kick model \NRSurRemnant~\cite{Varma:2019csw}
takes spins at $\trefmHundredM$ as input; therefore, if spin measurements are
available at $\frefTwentyHz$, they are first evolved using the \NRSur dynamics
from $\frefTwentyHz$ to $\trefmHundredM$ before computing the
kick~\cite{Varma:2019csw, Varma:2018aht}.  As discussed in our companion paper
\cite{Varma:2021csh}, this is equivalent to measuring the spins directly at
$\trefmHundredM$. Therefore, by construction, we get the same kick velocity for
individual events, independent of the reference point (modulo \NRSur spin
evolution errors, which are small compared to the model
errors~\cite{Varma:2019csw, Blackman:2017pcm}).

However, the same logic does not apply at the population level. Because the
orbital-plane spin angles are poorly constrained at $\frefTwentyHz$ for
individual events, this information can get diluted at the population level
(cf.~Fig.~\ref{fig:spins_ppd_20Hz}). Therefore, even if we use the spin
population at $\frefTwentyHz$ and evolve spins drawn from this population to
$\trefmHundredM$, the orbital-plane spin angle information is already lost. On
the other hand, at $\trefmHundredM$, the orbital-plane spin angles are better
constrained for both individual events~\cite{Varma:2021csh} and on the
population level (cf.~Fig.~\ref{fig:spins_ppd_20Hz}), and this information can
lead to improved kick population constraints. For this reason, we expect that
as more observations become available, the $\phi_1$ and $\delphi$ populations
will be significantly better constrained at $\trefmHundredM$, leading to better
kick population constraints as well.

\subsection{Mock population study}
\label{sec:app_mock_pop}
To test the fidelity of the von Mises model in recovering $\phi_1$ and
$\delphi$ populations, we conduct a mock population study. For each of our
\numEv events, we pick the maximum-likelihood posterior sample, but we rotate
the orbital-plane spin angles and set them to constant values at
$\trefmHundredM$ for all events. We construct two such populations, one with
$\phi_1=-\pi/2$ and $\delphi=\pm \pi$ and another with $\phi_1=\pi/2$ and
$\delphi=0$. We inject the corresponding signals in simulated detector noise
and recover the spin population using our hierarchical analysis.

We use the \NRSur waveform model for the injections as well as the parameter
inference (with the \textsc{LALInference} package~\cite{Veitch:2014wba}). The
signals are injected into Gaussian noise from a simulated LIGO-Virgo network at
design sensitivity; however, we rescale the injected distance such that the SNR
matches that of the observed event. Therefore, our mock populations
approximately mimic the parameters and detector sensitivity for these events.

Figure~\ref{fig:spins_ppd_inj} shows the results from our hierarchical analysis
on these mock populations. For both mock populations, the $\phi_1$ and
$\delphi$ distributions show a clear preference for the region near the
injected values.
In some cases, the peak locations are slightly offset from the injected values.
Such shifts away from the true value are consistent with statistical error due
to Gaussian noise.
Interestingly, the $\delphi=0$ population is better recovered than the
$\delphi=\pm\pi$ population. This suggests that it is easier to constrain
$\delphi$ for binaries with $\delphi=0$, in agreement with
Refs.~\cite{Gerosa:2014kta, Trifiro:2015zda}.

\subsection{Results using all 46 GWTC-2 events}
\label{sec:app_pop_allEv}

All results shown so far were restricted to the \numEv signals with
$M\gtrsim60M_{\odot}$ so that we can use the \NRSur model. In addition to
generating \NRSur posteriors for these \numEv events, Ref.~\cite{Varma:2021csh}
also used the \PhenomT~\cite{Estelles:2021gvs} waveform model to analyze all 46
GWTC-2 binary BH events at $\frefTwentyHz$.  We now repeat our analysis for all
46 binary BH events from GWTC-2, using the \PhenomT posteriors from
Ref.~\cite{Varma:2021csh} for the remaining 15 events (listed in Tab. II of
Ref.~\cite{Varma:2021csh}). We use spins at $\frefTwentyHz$ for all events, as
spins at $\trefmHundredM$ are not available for \PhenomT.
Figure~\ref{fig:spins_ppd_allEv} shows constraints on the $\delphi$ population
using using all 46 events. While we simultaneously model all spin degrees of
freedom, we only show the $\delphi$ population for simplicity. Similar to
Sec.~\ref{sec:app_pop_20Hz}, the $\phi_1$ population is not well constrained
when the spins are measured at $\frefTwentyHz$.

The left-panel of Fig.~\ref{fig:spins_ppd_allEv} shows the posteriors for the
$\mu_{\delphi}$ and $\sigma_{\delphi}$ parameters. Compared to
Fig.~\spinpopfignum, we now see that the 1D $\sigma_{\delphi}$ posterior
is distinguishable from the prior for both prior choices. In particular, there
is a preference for small widths, while the $\mu_{\delphi}$ distribution still
peaks at $\sim \pm \pi$. As shown in the right-panel of
Fig.~\ref{fig:spins_ppd_allEv}, this leads to a stronger peak near $\sim \pm
\pi$ in the $\delphi$ population, compared to Fig.~\spinpopfignum.

While this reinforces our results using only \NRSur, it is important to
consider that \PhenomT can have biases in recovering the orbital-plane spin
angles~\cite{Varma:2021csh}. In addition, Ref.~\cite{Varma:2021csh}
found significant differences between the $\delphi$ posteriors for \NRSur and
\PhenomT for GW190521~\cite{Abbott:2020tfl}. This suggests that a detailed
study of the impact of waveform systematics on the $\delphi$ population is
necessary. We leave this exploration to future work.

\end{document}